\documentclass[acus]{JAC2000}

\usepackage{graphicx}

\newcommand{\bi}{\begin{itemize}}
\newcommand{\ei}{\end{itemize}}
\newcommand{\im}{\item}
\newcommand{\rt}{\rightarrow}
\newcommand{\etal}{\it et al. \rm}

\begin{document}
\title{\flushright{M12}\\[15pt] \centering 
\boldmath  R Measurements at High $Q^2$}

\author{F. A. Harris, University of Hawaii,
Honolulu, HI 96822, USA}

\maketitle

\begin{abstract}

Previous measurements of $R=\sigma(e^+e^-
\rightarrow \mbox{hadrons}) /\sigma(e^+e^-\rightarrow\mu^+\mu^-)$ at
high $Q^2$ are reviewed.  Recent $R$ measurement results, including
those from the Beijing Spectrometer Experiment, are described.  The
present status of $R$ measurements and future
measurement possibilities are summarized.

\end{abstract}

\section{Introduction}

The QED running coupling constant evaluated at the $Z$ pole, 
$\alpha(M^2_{Z})$, and the anomalous 
magnetic moment of the muon, $a_{\mu}=(g-2)/2$, are two fundamental 
quantities that are used to test the Standard Model (SM). 
The dominant uncertainties in both $\alpha(M^2_{Z})$ and $a_{\mu}^{SM}$ are 
due to the effects of hadronic vacuum polarization, which cannot be 
reliably calculated in the low energy region.  Instead, with the 
application of dispersion relations,
experimentally measured $R$ values are used to determine the vacuum 
polarization, where $R$ is the lowest order cross section 
for $e^+e^-\rightarrow\gamma^*\rightarrow \mbox{hadrons}$
in units of the lowest-order QED cross section for
$e^+e^- \rightarrow \mu^+\mu^-$, namely
$R=\sigma(e^+e^- \rightarrow \mbox{hadrons})/\sigma(e^+e^-\rightarrow
\mu^+\mu^-)$, where 
$\sigma (e^+e^- \rightarrow \mu^+\mu^-) = \sigma^0_{\mu \mu}=
4\pi \alpha^2(0) / 3s$.  Improved precision for $\alpha(M^2_{Z})$ also
narrows the allowed range of the Higgs mass prediction using electro
weak loop corrections.   For much more detail on the importance of $R$
measurements, see references \cite{pepn,kuhn,simon}.  

The uncertainty in $\alpha(M^2_Z)$ is introduced when it is
extrapolated to the $Z$-pole. 

$$\alpha (q^2) = \frac{\alpha_0}{1 - \Delta \alpha (q^2)}, $$
where $\alpha_0$ is the  fine structure constant,
which is known very precisely, and $\Delta \alpha (q^2)$ is the
vacuum polarization term.
$$\Delta \alpha (q^2) = \Delta \alpha_l (q^2) + \Delta_{\rm had}^{(5)} \alpha (q^2) + \Delta_{\rm top} \alpha (q^2) $$

\noindent The leptonic vacuum polarization, $\Delta \alpha_l (q^2)$,
can be calculated theoretically, and the top contribution,
$\Delta_{\rm top} \alpha (q^2)$ is absorbed as a parameter in the SM fit.
The dominant uncertainty is then due to the effects of hadronic vacuum polarization,
$\Delta_{\rm had}^{(5)}$, which is calculated
with experimentally determined $R$ values 
using dispersion relations.

Here, $R$ measurements above the center-of-mass (cm) energy
corresponding to the $J/\psi$ are reviewed. $R$ measurements at lower
energy are described in ref.~\cite{simon,zhengguot}. Since the
theoretical
$a_{\mu}=(g-2)/2$ is more dependent on low energy than on high
energy $R$ values, we will concern ourselves here primarily with the effect
of $R$ measurements on $\alpha (q^2)$.
In the following sections, previous $R$ measurements will be summarized, 
recent measurements by
the Beijing Spectrometer Experiment (BES) and CLEO will be described, and the
current $R$ status will be reviewed.

\section{\boldmath Previous $R$ Measurements}

\begin{figure}[h]
\centering
\includegraphics*[width=85mm]{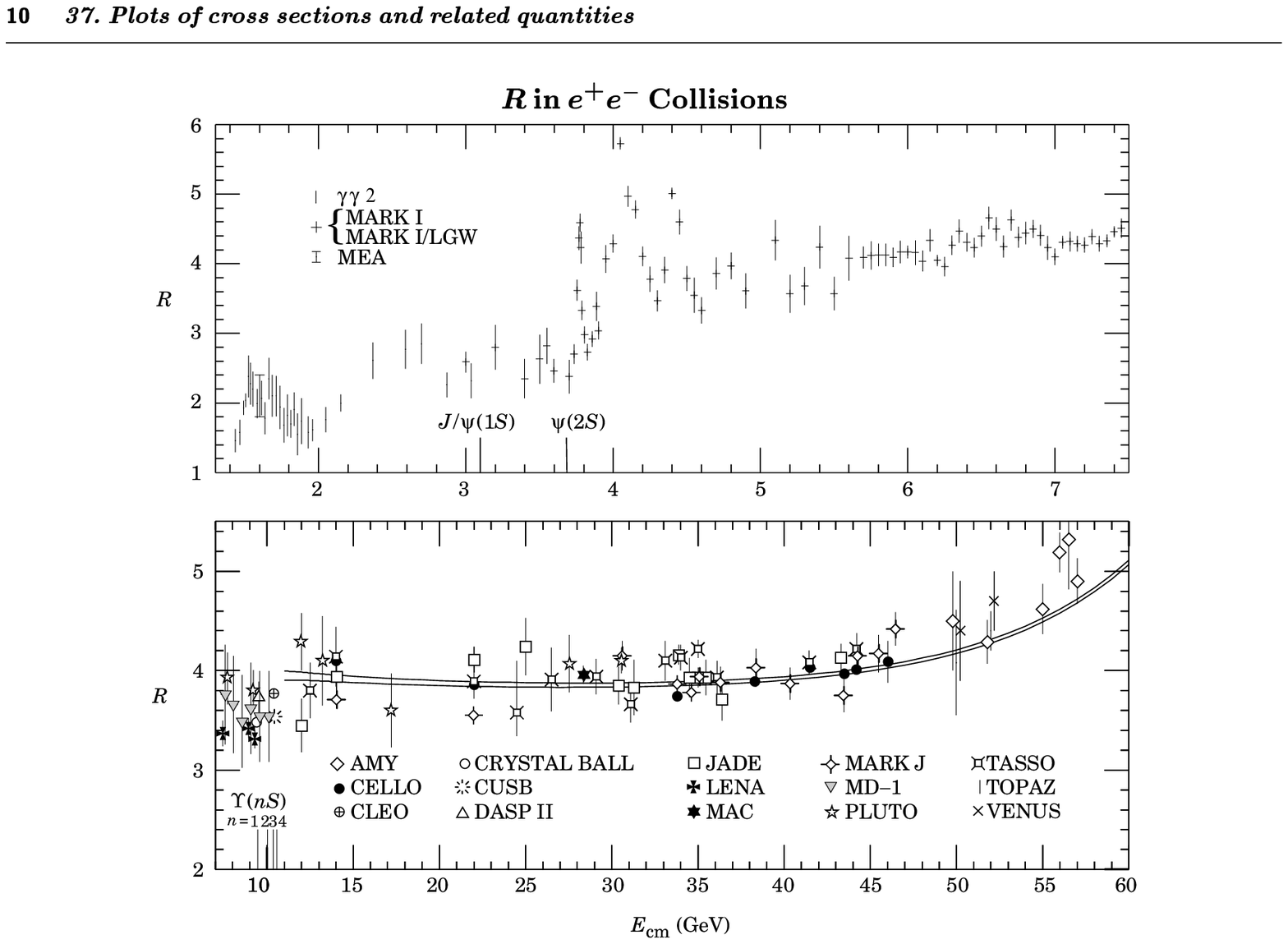}
\caption{Summary of $R$ values from PDG2000 \cite{pdg2000}.  Some
older measurements with even larger errors are not shown.  Systematic
normalization errors (5 - 20 \%) are not included on the points
shown.} \label{fig:pdg}
\end{figure}

Figure \ref{fig:pdg} shows the summary of $R$ measurements as
summarized by the Particle Data Group \cite{pdg2000}.  These are  
selected measurements; older measurements with even larger errors are not
shown.  Also 
systematic normalization errors (5 - 20 \%) are not included on the
points shown.  

We will be primarily interested in the center of mass (cm) energy region 
$ m_{J/\psi} < E_{\rm cm} < 12$ GeV.  
For higher energies, perturbative QCD
(PQCD) can be used to describe the behavior of $R$ as a function of
energy \cite{pietrzyk}.
Analyses that use data below $\sim 12$ GeV are ``data driven''
approaches \cite{pietrzyk};
``theory driven'' approaches use PQCD to go much lower in energy
\cite{davier}.
Although
less dependent on the quality of experimental data, the latter must
make additional theoretical assumptions.   

\begin{figure}[h]
\centering
\includegraphics*[width=80mm]{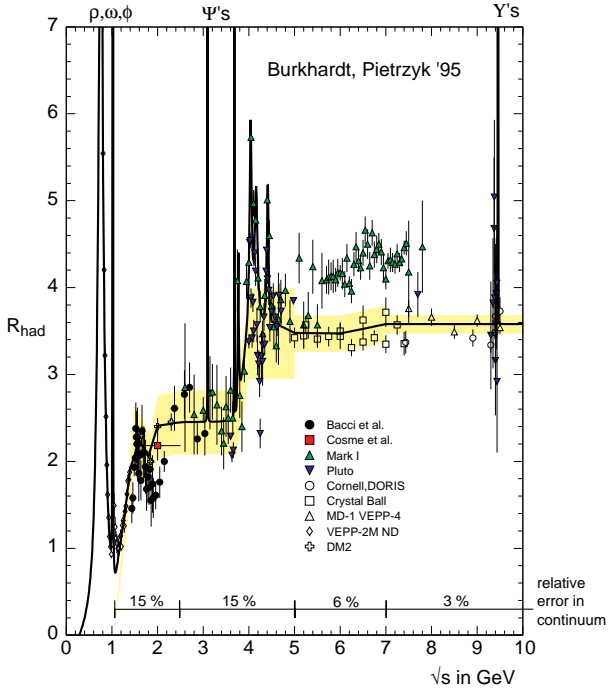}
\caption{Summary of 1995 $R$ values from ref. \cite{bolek95}. 
The uncertainties in the various energy regions are shown by the bands
and given at the bottom of the figure.} \label{fig:bolek95}
\end{figure}

Figure~\ref{fig:bolek95} shows a 1995 summary plot of the cm energy
region below 10 GeV \cite{bolek95}.  The $R$ values used as a function
of energy to evaluate the dispersion integral are indicated by the
smooth line through the points, and the $R$ value uncertainties
ascribed in the various energy regions are indicated by the bands.  In
the 1 to 5 GeV region, the $R$ uncertainty is taken to be 15 \%.  Note
that the Crystal Ball data \cite{crystalball} used in the 5 to 7 GeV
region is unpublished.

\section{\boldmath BES $R$ Measurement}

Recently a detailed $R$ scan was completed by the upgraded Beijing
Spectrometer (BES-II) Experiment \cite{zhengguo}.  The analysis is
described in some detail here to demonstrate the complexity of $R$
measurements. BESII is a general
purpose solenoidal detector located at the Beijing Electron Positron
Collider (BEPC), which is the only facility operating in the CM energy
range from 2 to 5 GeV. The luminosity at the $J/\psi$ is $\sim
5 \times 10^{30}$ cm$^{-2}$s$^{-1}$. BESII is described in detail elsewhere
\cite{besii}.

Experimentally, the value of $R$ is determined from the number of 
observed hadronic events, $N^{obs}_{had}$, by the relation
\begin{equation}
R=\frac{ N^{obs}_{had} - N_{bg} - \sum_{l}N_{ll} - N_{\gamma\gamma} }
{ \sigma^0_{\mu\mu} \cdot L \cdot \epsilon_{had} \cdot \epsilon_{trg}
\cdot (1+\delta)},
\end{equation}
where $N_{bg}$ is the number of beam-associated background events;
$\sum_{l}N_{ll},~(l=e,\mu,\tau)$ are the numbers
of lepton-pair events from one-photon processes and $N_{\gamma\gamma}$
the number of two-photon process
events that are misidentified as hadronic events;
$L$ is the integrated luminosity; $\delta$ is
the radiative correction; $\epsilon_{had}$ is the detection efficiency 
for hadronic events; and $\epsilon_{trg}$ is the trigger efficiency. 

In 1998 BES performed an initial measurement of $R$ using six scan
points: 2.6, 3.2, 3.4, 3.55, 4.6, and 5.0 GeV \cite{besr_1}.  At each point
separated beam runs were done for the study of beam gas background.
In 1999, BES measured 85 scan points with $\sim 1000$ hadronic events
per point.  Separated beam running was done at 24 energy points, while
single beam running was done at 7 points \cite{zhengguo}.

\subsection{Event Selection}

  The main sources of background for this measurement
are cosmic rays, lepton pair production, two-photon 
processes and single-beam associated processes.
Clear Bhabha events are first rejected.  Then the hadronic 
events are selected based on charged track information. 
Special attention is paid to two-prong events, where
cosmic ray and lepton pair backgrounds are especially severe,
and additional requirements are imposed to provide 
extra background rejection~\cite{besr_1}.     

An acceptable charged track must be in the
polar angle region $|\cos\theta|< 0.84$, have a good helix
fit, and not be clearly identified as an electron or muon.
The distance of closest approach to the beam axis must
be less than 2 cm in the transverse plane, and must occur
at a point along the beam axis for which $|z|<18$ cm.
In addition, the following
criteria must be satisfied: (i) $p < p_{beam} + 5 \times \sigma_p$,
where $p$ and $p_{beam}$ are the track and incident beam
momenta, respectively, and $\sigma_p$ is the momentum
uncertainty for a charged track for which $p = p_{beam}$;
(ii) $E < 0.6 E_{beam}$, where $E$ is the barrel shower counter (BSC)
energy associated
with the track, and $E_{beam}$ is the beam energy; (iii)
$2 < t < t_p + 5 \times \sigma_t$ (in ns.), where $t$ is the
measured time-of-flight for the track, $t_p$ is the
time-of-flight calculated assigning the proton mass to
the track, and  $\sigma_t$ is the resolution of the barrel time-of-flight system.

After track selection, event selection requires
the presence of at least two charged tracks, of which
at least one satisfies all of the criteria listed above.
In addition, the total energy deposited in the BSC ($E_{sum}$)
must be greater than $0.28 E_{beam}$, and the selected tracks
must not all point into the forward ($\cos \theta > 0$) 
or the backward ($\cos \theta < 0$) hemisphere.

For two-prong events, residual cosmic ray and lepton pair ($e^+ e^-$
and $\mu^+ \mu^-$) backgrounds are removed by requiring that the tracks
not be back-to-back, and that there be at least two isolated energy
clusters in the BSC with $E>100$~MeV that are at least 15$^{\circ}$ in
azimuth from the closest charged track. This last requirement rejects
radiative Bhabha events.

These requirements eliminate virtually all cosmic
rays and most of the lepton pair ($e^+ e^-$ and
$\mu^+ \mu^-$) events. The remaining
background contributions due to lepton pairs ($N_{ll}$), including
$\tau^+ \tau^-$ production above threshold,
and two-photon events ($N_{\gamma \gamma}$) are estimated
using Monte Carlo simulations and subtracted as indicated
in Eq. (1).

Cuts used for selecting hadronic events were varied in a 
wide range, e.g. $\cos \theta$ from 0.75 to 0.90, $E_{sum}$ from
0.24$E_{beam}$ to 0.32$E_{beam}$, to estimate the systematic 
error arising from the event selection, which turned out to be the dominant 
systematic error as indicated in Table~\ref{tab:error}.  

\subsection{Beam Associated Background}

The largest background is due to beam associated background, $ N_{bg}$.
To determine the level of single-beam-induced backgrounds,  
the same hadronic event selection criteria 
are applied to separated-beam data, and the number of 
separated-beam events, $N_{sep}$, surviving these criteria is 
obtained.  The number of beam-associated background events, 
$N_{bg}$, in the corresponding hadronic
event sample is given by $N_{bg}=f \times N_{sep}$, where $f$ is
the ratio of the products of the pressure at the
collision region and the integrated beam current for colliding 
and separated-beam runs.

\begin{figure}[!h]
\centering
\includegraphics*[width=85mm]{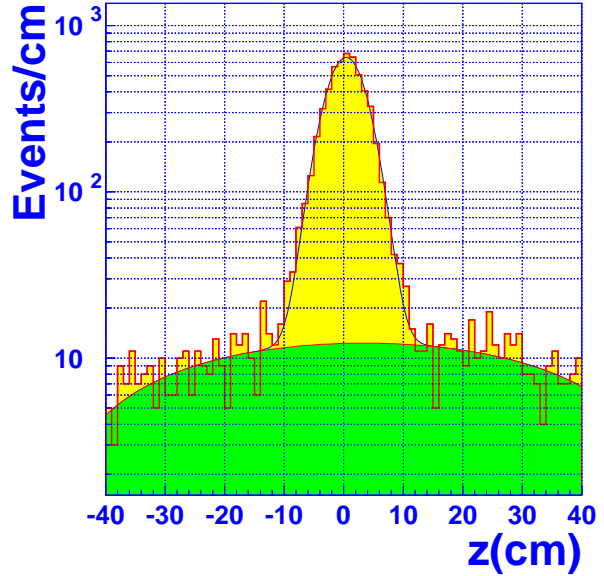}
\caption{Distribution of event vertices along the beam line ($z$).  A
gaussian is used to describe the real beam events, while a polynomial
of degree two is used to describe beam associated background.} \label{fig:evtvtx}
\end{figure}

\begin{figure}[!h]
\centering
\includegraphics*[width=85mm]{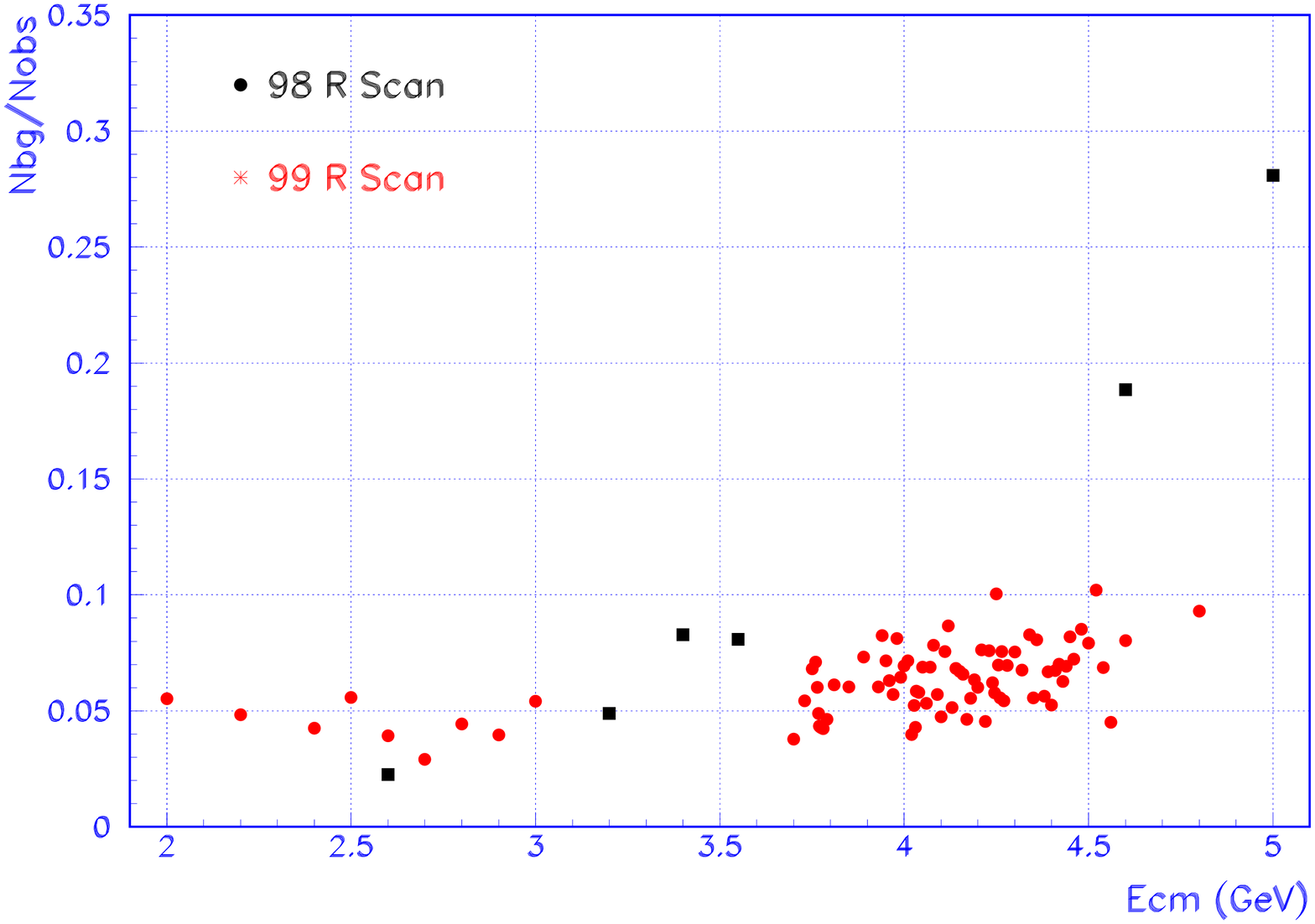}
\caption{Amount of beam associated background versus scan energy.
} \label{fig:backgrd}
\end{figure}

The beam-associated background can also be determined by fitting the
distribution of
event vertices along the beam direction with a Gaussian for real hadronic
events and a polynomial of degree two for the background, 
as shown in Fig.~\ref{fig:evtvtx}. This was the primary method used
for the '99 data.  Figure~\ref{fig:backgrd} shows the amount of beam associated
background versus the scan energy.
The differences between $R$ values obtained using these two methods to
determine the beam-associated background range between 0.3 and 2.3\%,
depending on the energy. These differences are included in the
systematic uncertainty.

\subsection{Hadron Efficiency}

JETSET, the Monte Carlo event generator that is commonly used to
simulate $e^+e^-\rt \mbox{hadrons}$, was not 
intended to be applicable to the low
energy region, especially that below 3 GeV. 
A special joint effort was made by the Lund 
group and the BES collaboration to develop the LUARLW generator,
which uses a formalism based on the 
Lund Model Area Law, but without the extreme-high-energy 
approximations used in JETSET's string fragmentation algorithm
~\cite{bo,huhm1}. 
The final states simulated in LUARLW are exclusive 
in contrast to JETSET, where they are inclusive. 
In addition, LUARLW uses fewer free 
parameters in the fragmentation function than JETSET.  
Above 3.77 GeV, 
the production of $D,~D^*,~D_s,$ and $D_s^*$
is included in the generator according to the Eichten 
Model~\cite{eichiten}.

\begin{figure}[!h]
\centering
\includegraphics*[width=82mm]{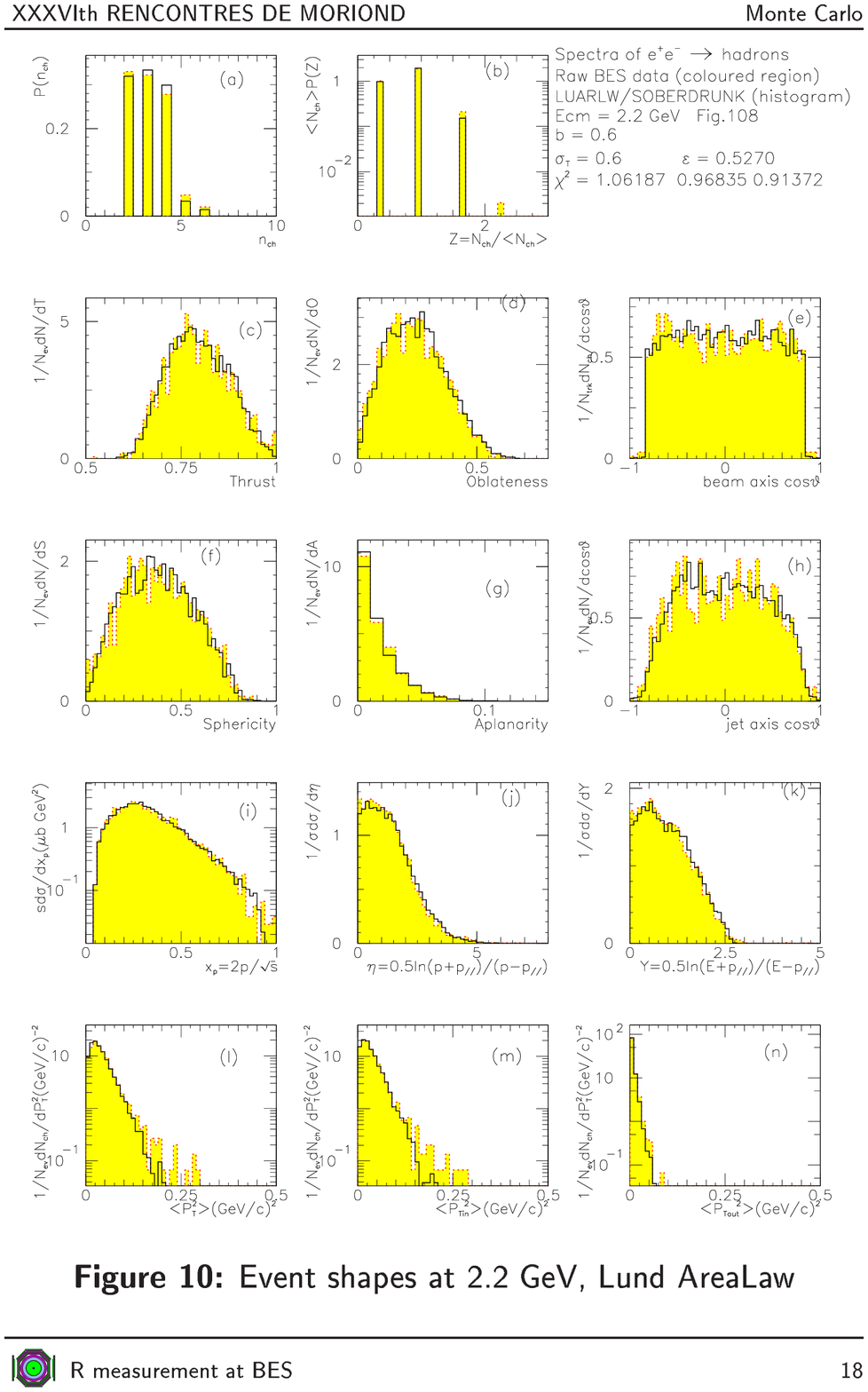}
\caption{Comparison of BES data (shaded histogram) and LUARLW Monte Carlo
data (black histogram) tuned for 2.2 GeV scan point.} \label{fig:lund}
\end{figure}

\begin{figure}[!h]
\centering
\includegraphics*[width=82mm]{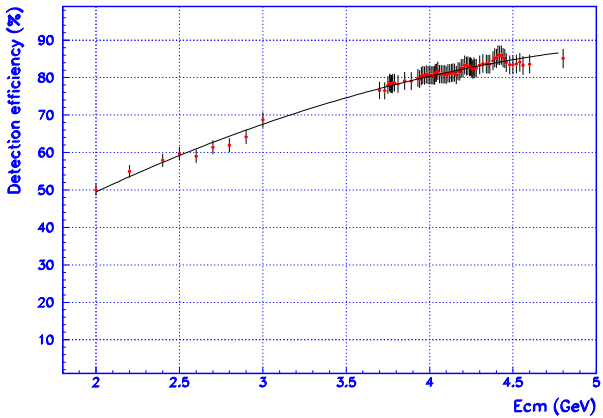}
\caption{The cm energy dependence of the detection efficiency
for hadronic events estimated using the LUARLW generator. The error 
bars are the total systematic errors.} \label{fig:eff_isr99}
\end{figure}

The parameters in LUARLW are tuned to reproduce 14 distributions of
kinematic variables over the entire energy region covered by the
scan~\cite{huhm1,huhm2}.  For example, the fits for parameters tuned at
$E_{\rm cm} = 2.2$ GeV are shown in Fig.~\ref{fig:lund}.
We find that one set of parameter values is required for the CM energy
region below open charm threshold, and that a second set
is required for higher energies.
In an alternative approach, the parameter values 
were tuned point-by-point throughout the
entire energy range.  
The detection efficiencies 
determined using individually tuned parameters are consistent with 
those determined with globally tuned parameters to within 2\%.   
This difference is included in the systematic errors. 
The detection efficiencies were also determined using JETSET74 for 
the energies above 3 GeV. 
The difference between the JETSET74 and LUARLW results is about 1\%,
and is also taken into account in estimating the systematic 
uncertainty.  Figure~\ref{fig:eff_isr99}~(a) shows the variation 
of the detection efficiency as a function of CM energy. 

\subsection{\boldmath Radiative Corrections: $(1 + \delta)$}

The  $(1 + \delta)$ term is necessary to
remove high order effects from $\sigma^{obs}_{had}$
\[ \sigma^{obs}_{had} = \sigma^o_{had} \cdot {\epsilon}_{had} \cdot
(1 + \delta)\]

\noindent Different schemes for the radiative corrections were compared
\cite{berends,fmartin,fadin,osterheld},
as reported in Ref. \cite{besr_1}. Below charm threshold, the four different
schemes agree with each other to within 1\%, while
above charm threshold, where resonances are
important, the agreement is within 1 to 3\%.  
The radiative correction used in this analysis is based on Ref.~\cite{osterheld},
and the differences with the other schemes are included in the
systematic error.

\subsection{Luminosity Measurement}

The integrated luminosity is determined from 
$$L = \frac{N_{obs}}{\sigma \: \epsilon \: \epsilon_{trg}} $$ using
Bhabha events.  The results obtained using
$\gamma \gamma$ and dimu events were
consistent.  The luminosity systematic error includes contributions
from cut
variations, background uncertainties, the cross section uncertainty, and
the efficiency uncertainty.  The luminosity systematic errors at some
selected scan energies are
shown in Table~\ref{tab:error}.

\subsection{Trigger Efficiency}
The trigger efficiency was
measured by comparing different trigger configurations
in special runs at the $J/\psi$.
$$\epsilon_{\rm Bhabha} = 99.96 \%$$
$$\epsilon_{\mu \mu} = 99.33 \% $$
$$\epsilon_{\rm had} = 99.76 \% $$
The 
error on the efficiencies is  0.5 \%.

\subsection{BES Results}

\begin{table*}[!hbtp]
\caption{Some values used in the determination of $R$ at a few
typical energy points.}
\begin{center}
\begin{tabular}{|ccccccccc|} \hline
$E_{cm}$ & $N_{had}^{obs}$ & $N_{ll}+$ & $L$  &
$\epsilon_{had}$ & $(1+\delta)$ & $R$  &
Stat.  & Sys.  \\
(GeV)  & & $N_{\gamma\gamma}$ & (nb$^{-1})$ & (\%) & &  & error & error \\
\hline
2.000 & 1155.4 & 19.5 &  47.3 & 49.50 & 1.024 & 2.18 & 0.07 & 0.18 \\
3.000 & 2055.4 & 24.3 & 135.9 & 67.55 & 1.038 & 2.21 & 0.05 & 0.11 \\
4.000 &  768.7 & 58.0 &  48.9 & 80.34 & 1.055 & 3.16 & 0.14 & 0.15 \\
4.800 & 1215.3 & 92.6 &  84.4 & 86.79 & 1.113 & 3.66 & 0.14 & 0.19 \\ \hline
\end{tabular}
\end{center}
\label{tab:value}
\end{table*}

\begin{table*}[hbtp]                                          
\caption{Contributions to systematic errors: hadronic selection,
luminosity determination, hadronic efficiency determination,
trigger efficiency, radiative corrections and total systematic error.
All errors are in percentages (\%).}
\begin{center}                        
\begin{tabular}{|ccccccc|} \hline
$E_{cm}$(GeV) & Had. sel. & $L$ & Had. eff. & Trig. & Rad. corr. & tot. \\
\hline
2.000 & 7.07 & 2.81 & 2.62 & 0.5 & 1.06 & 8.13 \\
3.000 & 3.30 & 2.30 & 2.66 & 0.5 & 1.32 & 5.02 \\
4.000 & 2.64 & 2.43 & 2.25 & 0.5 & 1.82 & 4.64 \\
4.800 & 3.58 & 1.74 & 3.05 & 0.5 & 1.02 & 5.14 \\ \hline
\end{tabular}
\end{center}
\label{tab:error}
\end{table*}

\begin{figure}[h]
\centering
\includegraphics*[width=85mm]{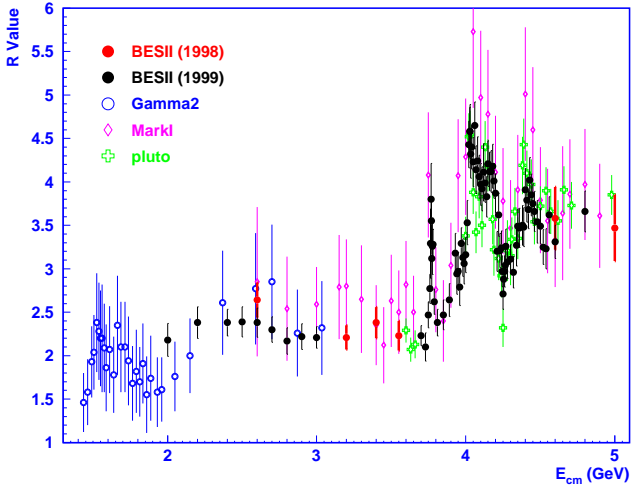}
\caption{A compilation of measurements of $R$ in the cm
energy range from 1.4 to 5 GeV.  In addition to the BES results
\cite{besr_1,besr_2}, results  from $\gamma \gamma2$,
Mark I, and Pluto \cite{gg2,mark1,pluto} are shown.} \label{fig:rvalue}
\end{figure}

The BES $R$ measurement results \cite{besr_1,besr_2} are shown in
Fig. ~\ref{fig:rvalue} along with the results from $\gamma \gamma2$,
Mark I, and Pluto \cite{gg2,mark1,pluto}.  Systematic uncertainties
are between 6 and 10 \% and are less than half of the previous
uncertainties.  The average uncertainty is 6.6 \%.
Tables~\ref{tab:value} and \ref{tab:error} list some of the values
used in the determination of $R$ and the contributions to the
uncertainty in the value of $R$ at a few typical energy points in the
scanned energy range, respectively.

\section{Recent CLEO Measurement}

\begin{figure}[h]
\centering
\includegraphics*[width=82mm]{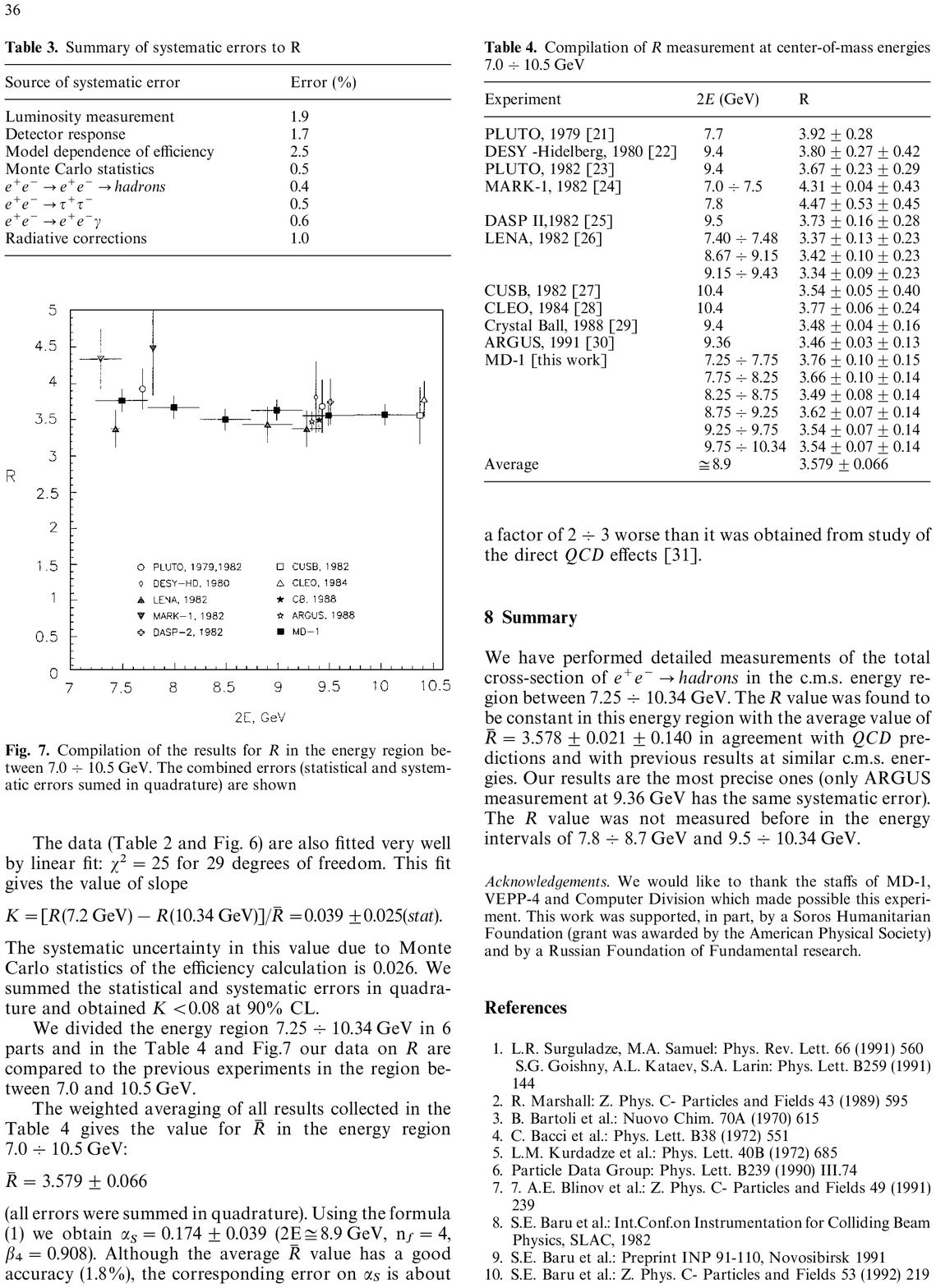}
\caption{$R$ measurements in the 7 to 10 GeV energy region.  Plot from
ref. \cite{blinov}.} \label{fig:10}
\end{figure}

The CLEO experiment 
measured $R$ at $E_{cm} = 10.52$ GeV, just below the $\Upsilon(4S)$,
and obtained
$R = 3.56 \pm 0.01 \pm 0.07 $ \cite{cleo}.
This 
very precise measurement (2 \%) obtained an error similar to that of
all previous results in this energy 
region combined,
$\bar{R} = 3.579 \pm 0.066$, corresponding to a 1.8 \% error \cite{blinov}.
Fig.~\ref{fig:10} shows the measurements in this region prior to CLEO \cite{blinov}.

The CLEO error is smaller than that obtained by BES.  A breakdown of the
error components is given in Table ~\ref{tab:compare}, along with the
corresponding errors at 3.0 GeV from BES.  Why does CLEO do better?  CLEO has
much better solid angle coverage and a higher detection efficiency, as
well as a state of the art detector.  They also ran with a higher
luminosity, 1.521 $\pm$ 0.015) fb$^{-1}$, and obtained 4 million
events. With a much bigger sample, both statistical and systematic
errors are reduced.

\section{\boldmath Current Status of $R$ Measurements}

\begin{figure}[h]
\centering
\includegraphics*[width=84mm]{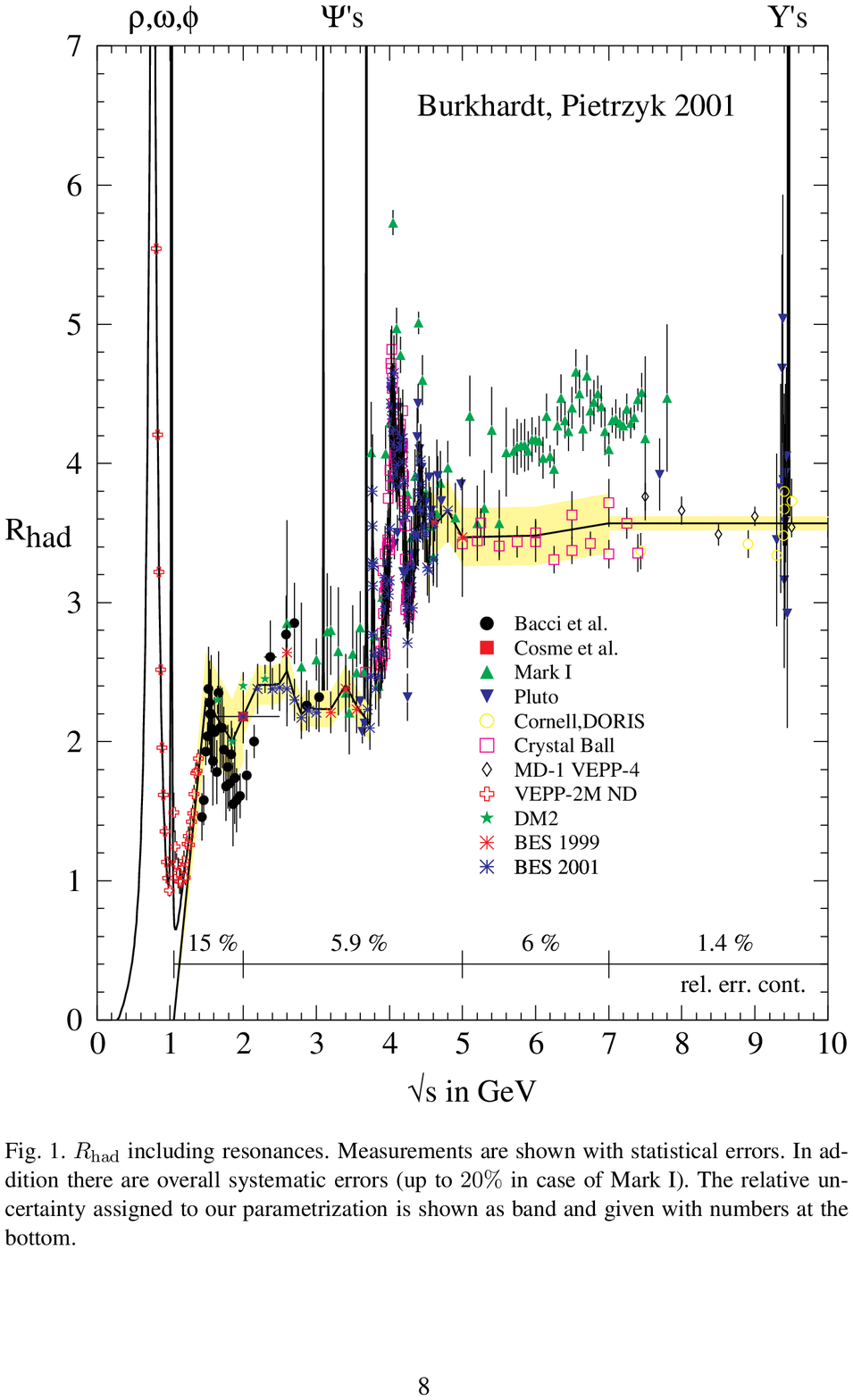}
\caption{Summary of current $R$ values from
ref.\cite{pietrzyk}. Compare with Fig. ~\ref{fig:bolek95}. } \label{fig:burkhardt01}
\end{figure}

\begin{figure}[h]
\centering
\includegraphics*[width=85mm]{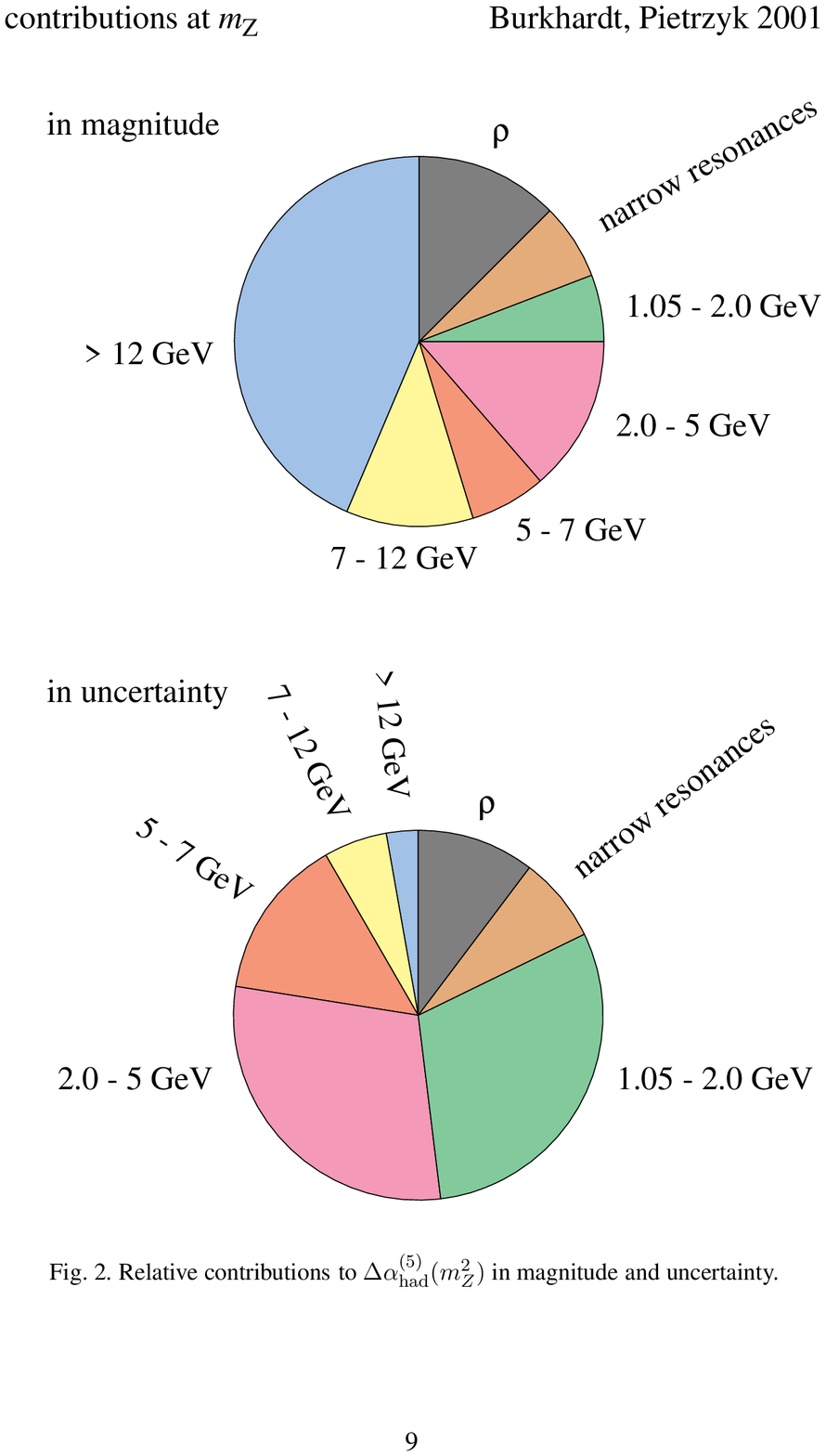}
\caption{Contributions to $\Delta^{(5)}_{\rm had}(m^2_Z)$ from the
various energy regions in magnitude and in uncertainty.  Figure from
ref. \cite{pietrzyk}. \label{fig:burkhardt02}}
\end{figure}

Burkhardt and  Pietrzyk have updated their analysis from 1995 \cite{bolek95}
with new results from CMD-2, CLEO, BES, and 3rd order QCD for
$E_{cm} > 12$ GeV \cite{pietrzyk}.
They find
$\alpha^{-1}(M^2_Z) = 128.936 \pm 0.046$,  and
$\Delta \alpha^{(5)}_{\rm had} = 0.02761 \pm 0.00036 $.
In 1995, $\Delta \alpha^{(5)}_{\rm had} = 0.0280 \pm 0.0007 $.
The improved experimental accuracy is primarily due to BES \cite{pietrzyk}.
Fig. ~\ref{fig:burkhardt01} shows their current summary of $R$
measurements below 10 GeV.  The error in the 2 to 5 GeV region is greatly reduced
because of the BES measurements. Figure ~\ref{fig:burkhardt02}
shows the breakdown of the error in $\Delta \alpha(M_Z^2)$ by energy region.
The biggest error contribution to $\Delta \alpha^{(5)}_{\rm had}$
still comes from the $1 < E_{cm} < 5$ GeV region!

\begin{table}[!h]
\caption{Comparison of $R$ systematic error contributions
 between BES (at 3.0 GeV) and CLEO (at 10.52 GeV).} 
\
\begin{center}
\begin {tabular}{|l|c|c|}\hline
~~Source~~~&~BES (\%)~ &  CLEO   \\
           & Error (\%)&  Error (\%)  \\ \hline
\label{tab:compare}
$N_{had}$          &        3.3  & -- \\
Backgd./Ev. Modeling &    --  & 0.7 \\
$L$                &        2.3  & 1.0 \\
$1 + \delta$       &        1.3  & -- \\
$\epsilon_{had}$   &        2.7  & -- \\ 
$\epsilon_{had} \times (1 + \delta)$ & -- & 1.0 \\  \hline

Sys.               &        5.0  & 1.8 \\
Stat.              &        2.5  & 0.3 \\ \hline
Total        &   5.6  & 2.0 \\ 
\hline
\end{tabular}
\end{center}
\end{table}

\section*{\boldmath Future $R$ Measurements}

 J. H. K\"uhn has used the new BES results to test PQCD,
but calls for even better precision (2 \%) \cite{kuhn}.
Certainly, the precision of $R$ in the 1 - 2 GeV cm region must be improved.
This is extremely important to improve the precision of both $a_{\mu}$ and $\alpha (M_Z^2)$.
The KLOE and PEP-N  experiments are candidates for measuring $R$ in
this region.

As shown in Fig.~\ref{fig:burkhardt02}, the  2 - 5 GeV region is also
very important in the
determination of  $\Delta \alpha (M_Z^2)$.  The candidates for $R$
measurements in this region are BES,
PEP-N, and CLEO-C.  Also improved measurements in this region are
important to 
clarify the structure in the charm resonance region.
The B Factories may also contribute to both energy regions using ISR
events \cite{xinchou,solodov}.  

The 5 -7 GeV region is also very important.  As shown in
Fig.~\ref{fig:burkhardt01}, the $R$ values used by theorists in this
region are the unpublished Crystal Ball values \cite{crystalball}.  CLEO-C could measure
points in this region.  

 What is needed to improve $R$ value precision?
\begin{enumerate}
\im High luminosity $\rt$ large sample
\im Good solid angle coverage
\im Excellent detector with good particle identification
\im Radiative correction to better than 1 \%
\im More effort on the event generator (LUARLW)
\im Large sample for tuning the generator
\im Measure exclusive channels at low energy\\
\end{enumerate}

\section{Energy reach of PEP-N}

Although very difficult, extending the energy range
to above the mass of the $\psi(2S)$ should be considered.
Currently the world's largest sample ( $\sim$4 M events) is that of
the BES experiment.
Some very good physics becomes
accessible in this region. $\tau \tau$ production at threshold can be
measured to improve the precision of the $\tau$ mass.
Approximately one quarter of  $\psi(2S)$ decays are to $\chi_c$ and
$\eta_c$.  Running on the  $\psi(2S)$ is a good place to do  $\chi_c$
and $\eta_c$ physics, as well as to  study $\psi(2S)$ hadronic decays.
The study of the hadronic decays might provide a solution to the well
known  $\rho \pi$ puzzle \cite{rhopi}.  Being able to compare  $J/\psi$ and
$\psi(2S)$ decays is also very beneficial in understanding the final states
in these decays.
Finally it will be very useful to have some overlap between PEP-N and
CLEO-C as a check of systematics in measuring $R$ in this region.

\section{Summary}

Although the precision of $R$ measurements has improved,
better $R$ precision is still needed to improve the precision of the
theoretical standard model values of 
$a_{\mu}$ and $\alpha (m^2_Z)$ and to test PQCD.
There are 
many interesting possibilities for future improvements
of $R$ values from KLOE, PEP-N, and CLEO-C, as well from the B
factories using ISR events.

\section{Acknowledgements}
I would like to thank the organizers of the {\em $e^+ e^-$ Physics at
Intermediate Energies Workshop} for their hospitality.  I also want to
thank Bolek Pietrzyk for many of the plots used in this paper and my
friend and collaborator Zhengguo Zhao, who is the leader of the
R-Group, for his help on this paper.  I also want to acknowledge the
efforts of the BES R-group, the Computing Group at IHEP, the BEPC
staff, and the members of the BES collaboration.



\begin{thebibliography}{9}

\bibitem{pepn} W. Marciano, ``General Talk on Hadronic Corrections
in the SM'',
{\em The $e^+ e^-$ Physics at
Intermediate Energies Workshop}, M06.

\bibitem{kuhn} J. H. K\"uhn, ``Hadronic Corrections to $\alpha(M_Z^2)$ and $g_{\mu}$,
Theory and Experimental Data'', {\em The $e^+ e^-$ Physics at
Intermediate Energies Workshop}, M08.

\bibitem{simon} S. Eidelman, ``Status of CVC Tests from
Electron-Positron Annihilation and $\tau$ Decay'', {\em The $e^+ e^-$ Physics at
Intermediate Energies Workshop}, M07.

\bibitem{zhengguot} Z. Zhengguo, ``Results and Future Plans from BEPC'',  {\em The $e^+ e^-$ Physics at
Intermediate Energies Workshop}, M10.


\bibitem{pdg2000}
Particle Data Group, D.E. Groom {\it et al.}, Eur. Phys. J.
C15, 1 (2000).

\bibitem{pietrzyk}
H. Burkhardt and B. Pietrzyk, accepted by
Phys. Lett. B, LAPP-EXP 2001-03 (Feb. 2001).


\bibitem{crystalball} C. Edwards {\it et al.},
SLAC PUB 5160 (1990).

\bibitem{davier} M. Davier and A. Hoecker, Phys Lett. {\bf B419}, 419 (1998).

\bibitem{bolek95} H. Burkhardt and B. Pietrzyk, Phys. Lett. {\bf
B356}, 398 (1995).

\bibitem{zhengguo} Z.G. Zhao, International Journal of Modern Physics A15,
3739 (2000).  

\bibitem{besii}J.Z. Bai {\it et al.}, (BES Collab.),
Nucl. Instr. Methods {\bf A458}, 627 (2001).

\bibitem{besr_1} J. Z. Bai {\it et al.}, (BES Collab.),
Phys. Rev. Lett. {\bf 84}, 594 (2000).

\bibitem{bo} B. Andersson and Haiming Hu, ``Few-body States in Lund String
Fragmentation Model'', hep-ph/9910285.

\bibitem{huhm1} For a more complete description of LUARLW, see
Haiming Hu, ``Hadron Production at Intermediate Energies'', {\em The $e^+ e^-$ Physics at
Intermediate Energies Workshop}, T24.

\bibitem{eichiten}E. Eichten {\it et al.}, Phys. Rev. {\bf D21}, 203 (1980).

\bibitem{huhm2} Haiming Hu {\it et al.}, "The Application of A New
Generator Based on Lund Area Law to the $R$ Scan in 2-5 GeV Center-of-mass
Energy Region", Accepted by "High Energy Physics and Nuclear Physics
(China)".  

\bibitem{berends}F.A. Berends and R. Kleiss, Nucl. Phys. {\bf B178},
141 (1981).

\bibitem{fmartin} G. Bonneau and F. Martin, Nucl. Phys. {\bf B27}, 387 (1971).

\bibitem{fadin} E. A. Kuraev and V.S. Fadin, {\em Sov. J. Nucl.
Phys.} {\bf41}, 3 (1985). 

\bibitem{osterheld} A. Osterheld {\it et al.}, SLAC-PUB-4160, 1986. (T/E)

\bibitem{besr_2}  J. Z. Bai {\it et al.}, (BES Collab.), submitted to
PRL, hep-ex/102003.

\bibitem{gg2} C. Bacci {\it et al.}, ($\gamma \gamma2$ Collab.),
Phys. Lett. {\bf B86}, 234 (1979).


\bibitem{mark1} J. L. Siegrist {\it et al.}, (Mark I Collab.),
Phys. Rev. {\bf D26}, 969 (1982).         

\bibitem{pluto} L. Criegee and G. Knies, (Pluto Collab.),
Phys. Rep. {\bf 83}, 151 (1982);\\
Ch. Berger {\it et al.}, Phys. Lett. {\bf B81}, 410 (1979).

\bibitem{cleo}
R. Ammar \etal, CLEO Collab., Phys. Rev. D57, 1350 (1998).

\bibitem{blinov}
A. E. Blinov \etal, Z. Phys. C 70, 31 (1996).

\bibitem{xinchou}
T. Benninger, X.C. Lou, and W. M. Dunwoodie, ``Physics with ISR 
EVENTS at B Factory Experiments''

\bibitem{solodov}  E. Solodov, Study of the 1.5 - 3 GeV region using
ISR with BaBar'', {\em The $e^+ e^-$ Physics at
Intermediate Energies Workshop}, T03.

\bibitem{rhopi} M. Suzuki, Phys. Rev. {\bf D63}, 054021 (2001).

\end{thebibliography}
\end{document}